\documentclass[twocolumn,showpacs,superscriptaddress,amsmath,amssymb,aps,pre]{revtex4}
\usepackage{graphicx}% Include figure files
\usepackage{dcolumn}% Align table columns on decimal point
\usepackage{bm}% bold math
\usepackage{wrapfig,subfigure}
\usepackage{amssymb}
\usepackage{color}
\def\tr{t_{\rm r}}

\def\Tr{{\rm Tr}}

\newcommand{\qb}{{\bf q}}
\newcommand{\pb}{{\bf p}}
\newcommand{\fb}{{\bf f}}

\newcommand{\xib}{\bm\xi}
\newcommand{\ph}{p_h({\bf q},t|{\bf q}_f,t_f)}
\def\p12{p_{12}({\bf q},t)}

\begin{document}

\title{Switching path distribution in multi-dimensional systems}% Force line breaks with \\

\author{H. B. Chan}
\email{hochan@phys.ufl.edu}
\affiliation{Department of Physics, University of Florida, Gainesville, Florida 32611}
\author{M. I. Dykman}
\email{dykman@pa.msu.edu}
\affiliation{Department of Physics and Astronomy, Michigan State University, East Lansing, Michigan 48823}
\author{C. Stambaugh}
\affiliation{Department of Physics, University of Florida, Gainesville, Florida 32611}

\begin{abstract}
We explore the distribution of paths followed in fluctuation-induced switching between coexisting stable states. We introduce a quantitative characteristic of the path distribution in phase space that does not require a priori knowledge of system dynamics. The theory of the distribution is developed and its direct measurement is performed in a micromechanical oscillator driven into parametric resonance. The experimental and theoretical results on the shape and position of the distribution are in excellent agreement, with no adjustable parameters. In addition, the experiment provides the first demonstration of the lack of time-reversal symmetry in switching of systems far from thermal equilibrium. The results open the possibility of efficient control of the switching probability based on the measured narrow path distribution.
\end{abstract}

\maketitle

%%%%%%%%%%%%%%%%%%%%%%%%%%%%%%%%%%%%%%%%%%%%%%%%%%%%%%%%%%
%Article
%%%%%%%%%%%%%%%%%%%%%%%%%%%%%%%%%%%%%%%%%%%%%%%%%%%%%%%%%%
\section{INTRODUCTION}

Fluctuation phenomena in systems with multiple stable states have long been a topic of intense research interest. When the fluctuation intensity is small, for most of the time the system fluctuates about one of the stable states. Switching between the states require large fluctuations that would allow the system to overcome the activation barrier. Such large fluctuations are rare. However, they lead to large changes in the system behavior. Fluctuational switching between coexisting states plays a crucial role in a variety of systems and phenomena including nanomagnets \cite{Wernsdorfer1997}, Josephson junctions \cite{Fulton1974}, protein folding \cite{Wales_book2003}, and chemical reactions.

A detailed theory of switching rates was first developed by Kramers \cite{Kramers1940}. The analysis referred to systems close to thermal equilibrium, where the switching rate is determined by the free energy barrier between the states. The concept of this barrier is well understood and the barrier height has been found for many models. However, in recent years much attention has been given to switching in systems far from thermal equilibrium. Examples include electrons \cite{Lapidus1999} and atoms \cite{Gommers2005,Kim2006} in modulated traps, rf-driven Josephson junctions \cite{Siddiqi2004,Lupascu2007} and nano- and micromechanical resonators \cite{Aldridge2005,Stambaugh2006,Almog2007}. Apart from fundamental interest, switching in modulated systems is important for many applications, quantum measurements being an example \cite{Siddiqi2004,Lupascu2007,Schwab2005a,Katz2007}. Nonequilibrium systems generally lack detailed balance, and the switching rates may not be found by a simple extension of the Kramers approach.

A basic, although somewhat counterintuitive, physical feature of large infrequent fluctuations leading to switching is that, in such fluctuations, the system is most likely to move along a certain path in its phase space. This path is known as the most probable switching path (MPSP). For low fluctuation intensity, the MPSP is obtained by the solution of a variational problem. This problems also determines the switching activation barrier \cite{Freidlin_book,Dykman1979a,Graham1984,Luciani1987,Bray1989,Dykman1990,Maier1993,Herwaarden1995,Lehmann2000a,Kraut2004,Elgart2004}.  Despite its central role for the understanding of fluctuation-induced switching and switching rates, the idea of the MPSP has not been tested experimentally in multivariable systems. The question of how the paths followed in switching are distributed in phase space has not been asked either.

Perhaps closest to addressing the above issues was the experiment on dropout events in a semiconductor laser with optical feedback \cite{Hales2000}. In this experiment the switching path distribution in space and time was measured and calculated. However, the system was characterized by only one dynamic variable, and thus all paths lie on one line in phase space. In another effort, electronic circuit simulations \cite{Luchinsky1997} compare the distribution of fluctuational paths to and relaxational paths from a certain point within one basin of attraction; the data refer to the situation where switching does not occur. The methods \cite{Hales2000,Luchinsky1997} do not apply to the switching path distribution in multivariable systems, as explained  below.

In the present paper we introduce the concept of switching path distribution in phase space and the quantity that describes this distribution, calculate this quantity, and report the direct observation of the tube of switching paths. A brief account of the results was published in Ref.~\onlinecite{Chan2008}. The experimental and theoretical results on the shape and position of the path distribution are in excellent agreement, with no adjustable parameters. The results open the possibility of efficient control of the switching probability based on the measured narrow path distribution.

In Sec.~\ref{sec:preview} we provide the qualitative picture of switching and give a preview of the central theoretical and experimental results. Sec.~\ref{sec:theory} presents a theory of the switching probability distribution in the basins of attraction to the initially occupied and initially empty stable states as well as some simple results for systems with detailed balance. In Sec.~\ref{sec:system} the system used in the experiment, a micromechanical torsional oscillator, is described and quantitatively characterized. Section~\ref{sec:experiment} presents the results of the experimental studies of the switching path distribution for the micromechanical oscillator, with the coexisting stable states being the states of parametrically excited nonlinear vibrations. Generic features of the distribution are discussed, and the lack of time-reversal symmetry in switching of systems far from thermal equilibrium is demonstrated for the first time. Section~\ref{sec:conclusions} contains concluding remarks.

\section{QUALITATIVE PICTURE AND PREVIEW OF THE RESULTS}
\label{sec:preview}

\begin{figure}[h]
\includegraphics[width=3.0in,angle=0]{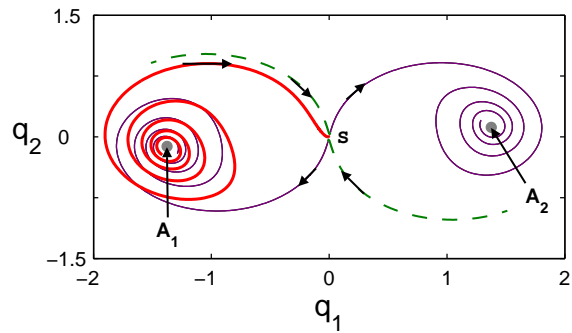}
\caption{(Color online) Phase portrait of a two-variable system with two stable states $A_1$ and $A_2$. The saddle point ${\cal S}$ lies on the separatrix that separates the corresponding basins of attraction. The thin solid lines show the downhill deterministic trajectories from the saddle to the attractors. A portion of the separatrix near the saddle point is shown as the dashed line. The thick solid line shows the most probable trajectory that the system follows in a fluctuation from $A_1$ to the saddle. The MPSP from $A_1$ to $A_2$ is comprised by this uphill trajectory and the downhill trajectory from ${\cal S}$ to $A_2$. The plot refers to the system studied experimentally, see Sec.~\ref{sec:experiment}.}
\label{fig:phase_portrait}
\end{figure}

We consider a bistable system with several dynamical variables  $\qb = (q_1,...,q_N)$. The stable states $A_1$  and $A_2$ are located at $\qb_{A_1}$ and $\qb_{A_2}$  respectively. A sketch of the phase portrait for the case of two variables is shown in Fig.~\ref{fig:phase_portrait}. For low fluctuation intensity, the physical picture of switching is as follows. The system prepared initially in the basin of attraction of state $A_1$, for example, will approach $\qb_{A_1}$ over the characteristic relaxation time $t_r$ and will then fluctuate about $\qb_{A_1}$. We assume the fluctuation intensity to be small. This means that the typical amplitude of fluctuations about the attractor (the characteristic diffusion length) $l_D$ is small compared to the minimal distance $\Delta q$ between the attractors and from the attractors to the saddle point $\qb_{\cal S}$.

Even though fluctuations are small on average, occasionally there occur large fluctuations, including those leading to switching between the states. The switching rate $W_{12}$ from state $A_1$ to $A_2$ is much less than the reciprocal relaxation time $t_r^{-1}$, that is, the system fluctuates about $A_1$ for a long time, on the scale of $t_r$, before a transition to $A_2$ occurs. In the transition the system most likely moves first from the vicinity of $\qb_{A_1}$ to the vicinity of $\qb_{\cal S}$. Its trajectory is expected to be close to the one for which the probability of the appropriate large rare fluctuation is maximal. The corresponding trajectory is illustrated in Fig.~\ref{fig:phase_portrait}. From the vicinity of $\qb_S$ the system moves to state $A_2$ close to the deterministic fluctuation-free trajectory. These two trajectories comprise the MPSP.

For brevity, we call the sections of the MPSP from $\qb_{A_1}$ to $\qb_{\cal S}$ and from $\qb_{\cal S}$ to $\qb_{A_2}$ the downhill and uphill trajectories, respectively. The terms would literally apply to a Brownian particle in a potential well, with $A_{1,2}$ corresponding to the minima of the potential and ${\cal S}$ to the barrier top.

%%%%%%%%%%%%%
%%%%%%%%%%%%%

We characterize the switching path distribution by the probability density for the system to pass through a point $\qb$ on its way from $A_1$ to $A_2$,
\begin{equation}
\label{eq:1}
p_{12}(\qb,t)=\int_{\Omega_2}d\qb_f\rho(\qb_f,t_f;\qb,t|\qb_0,t_0).
\end{equation}
%(1)
Here, the integrand is the three-time conditional probability density for the system to be at points $\qb_f$  and $\qb$  at times $t_f$  and $t$, respectively, given that it was at $\qb_0$  at time $t_0$. The point $\qb_0$  lies within distance $\sim l_D$  of $\qb_{A_1}$  and is otherwise arbitrary. Integration with respect to $\qb_f$  goes over the range $\Omega_2$  of small fluctuations about $\qb_{A_2}$; the typical linear size of this range is $l_D$.

We call $\p12$ the switching probability distribution. Of utmost interest is to study this distribution in the time range
\begin{equation}
\label{eq:time_range}
W_{12}^{-1},W_{21}^{-1} \gg t_f-t,t-t_0 \gg \tilde{t_r}.
\end{equation}
Here, $\tilde{t_r}$ is the Suzuki time \cite{Suzuki1977}. It differs from $t_r$ by a logarithmic factor $\sim\log[\Delta q/l_D]$. This factor arises because of the motion slowing down near the saddle point. The time $\tilde{t_r}$ is much smaller than the reciprocal switching rates, and the smaller the fluctuation intensity the stronger the difference, because the dependence of $W_{ij}$ on the fluctuation intensity is of the activation type. If the noise causing fluctuations has a finite correlation time, $\tilde{t_r}$ is the maximum of the Suzuki time and the noise correlation time.

For $t-t_0 \gg t_r$, by time $t$ the system has already ``forgotten" the initial position $\qb_0$. Therefore the distribution $\rho(\qb_f,t_f;\qb,t|\qb_0,t_0)$, and thus $p_{12}$, are independent of $\qb_0,t_0$. On the other hand, if the system is on its way from $A_1$ to $A_2$ and at time $t$ is in a state $\qb$ far from the attractors, it will most likely reach the vicinity of $A_2$ over time $\lesssim \tilde{t_r}$ and will then fluctuate about $\qb_{A_2}$. This will happen well before the time $t_f$ at which the system is observed near $A_2$, and therefore $p_{12}$ is independent of $t_f$.

It is clear from the above arguments that, in the time range (\ref{eq:time_range}), the distribution $\p12$ for $\bf q$ far from the attractors is formed by switching trajectories emanating from the vicinity of $A_1$. It is gives the probability density for these trajectories to pass through a given point $\qb$ at time $t$. In other terms, the distribution $\p12$ is formed by the probability current from $A_1$ to $A_2$ and is determined by the current density.

\subsection{The shape of the switching probability distribution}

We show in Sec.~\ref{sec:theory} that $\p12$ peaks on the MPSP. The peak is Gaussian transverse to the MPSP for $ \vert\qb- \qb_{A_{1,2}}\vert,\vert\qb - \qb_{\cal S}\vert \gg l_D$,
\begin{equation}
 \label{eq:2}
 p_{12}(\qb,t)= W_{12}v^{-1}(\xi_{\parallel})Z^{-1}\exp\left({-\frac{1}{2}
\xib_{\perp}\widehat{Q}\xib_{\perp}}\right),
 \end{equation}
%(2)
where $\xi_{\parallel}$  and ${\xib_{\perp}}$  are coordinates along and transverse to the MPSP, and $v(\xi_{\parallel})$  is the velocity along the MPSP. The matrix elements of matrix $\hat{Q}=\hat{Q}(\xi_{\parallel})$  are  $\propto l_D^{-2}$, and  $Z=[(2\pi)^{N-1}/\det\hat{Q}]^{1/2}$. It follows from Eq.~(\ref{eq:2}) that the overall probability flux along the MPSP is equal to the switching rate,
\[\int d\xib_{\perp}\p12 v(\xi_{\parallel})=W_{12}.\]

\begin{figure}[h]
\includegraphics[width=3.0in]{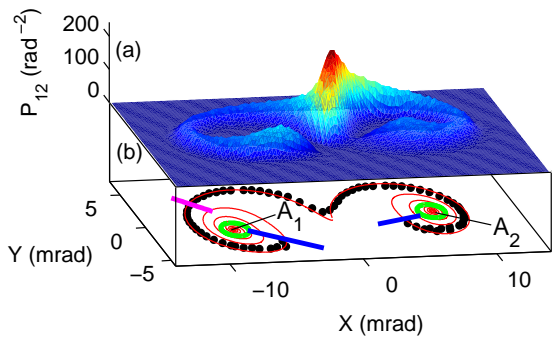}
\caption{(Color) (a) Switching probability distribution in a parametrically driven microelectromechanical oscillator. The probability distribution $p_{12}(X,Y)$ is measured for switching out of state $A_1$ into state $A_2$. (b) The peak locations of the distribution are plotted as black circles and the theoretical most probable switching path is indicated by the red line. All trajectories originate from within the green circle in the vicinity of $A_1$ and later arrive at the green circle around $A_2$. The portion of the distribution outside the blue lines is omitted. }
\label{fig:1}
\end{figure}

We have observed a narrow peak of the switching path distribution in experiment. The results are shown in Fig.~\ref{fig:1}. They were obtained using a micro-electro-mechanical torsional oscillator described in Sec.~\ref{sec:system}. The path distribution displays a sharp ridge. We demonstrate that the cross-section of the ridge has Gaussian shape. As seen from Fig.~\ref{fig:1}, the maximum of the ridge lies on the MPSP which was calculated for the studied system.

%%%%%%%%%%%%%
%%%%%%%%%%%%%

Equation (\ref{eq:2}) is written for a generally nonequilibrium system, but the system is assumed to be stationary. In the neglect of fluctuations its motion is described by equations with time-independent coefficients. In this case $\p12$ is independent of time $t$. A different situation may occur in periodically modulated systems. In such systems, attractors are periodic functions of time. If the typical relaxation time is smaller than or of the order of the modulation period, generally there is one MPSP per period. Then $\p12$ is also a periodic function of time. We will not consider this case in the present paper.

\subsection{Comparison with the prehistory distribution}
\label{subsec:prehistory}

The distribution of fluctuational paths was studied earlier in the context of the ``prehistory problem" \cite{Dykman1992d}. In this formulation one is interested in the paths to a certain state $\qb_f$ that is far from the initially occupied attractor. The distribution of these paths $p_h$ is given by the probability density to have passed a point $\qb$ at time $t$ given that the system is found at $\qb_f$ at a later time $t_f$ whereas initially at time $t_0$ it was at point $\qb_0$ near attractor $A_1$,
\begin{equation}
\label{eq:standard_prehistory}
\ph=\frac{\rho(\qb_f,t_f;\qb,t|\qb_0,t_0)}{\rho(\qb_f,t_f;\qb_0,t_0)}.
\end{equation}

The prehistory distribution (\ref{eq:standard_prehistory}) and its generalizations were analyzed in a number of papers \cite{Hales2000,Luchinsky1997,Morillo1997,Vugmeister1997a,Mannella1999}. However, the problem of paths that lead to switching between the states was addressed only for a stationary system with one dynamical variable \cite{Hales2000}. In this case, the system must pass through all the intermediate points between the two states during a switch. For systems with more than one dynamical variable, the formulation \cite{Hales2000,Luchinsky1997,Morillo1997} no longer applies because
it cannot be known in advance through what points the system will pass in switching. The aforementioned formulation does not work even for one-variable periodically modulated systems, since the distribution (\ref{eq:standard_prehistory}) depends not only on the position of point $q_f$, but also on the time $t_f$ when this point is reached \cite{Ryvkine2006}.

In contrast, the distribution $\p12$ is defined in such a way that it is independent of the final point $\qb_f$ and of the time $t_f$ of reaching it. The definition does not impose any constraint on paths except that they lead to switching between the attractors. Therefore the introduction of the function $\p12$ is essential in studying the switching path distribution for multivariable systems.

\section{Theory of the switching path distribution}
\label{sec:theory}

\subsection{The model of a fluctuating system}
\label{subsec:white_noise}

We derive Eq.~(2) for a system described by the Langevin equation of motion
\begin{eqnarray}
\label{eq:Langevin}
 \dot \qb = {\bf K}(\qb) + \fb(t), &
 \langle f_n(t) f_m(t')\rangle = 2D\delta_{nm}\delta(t-t').
 \end{eqnarray}
%(3)
Here, the vector ${\bf K}$ determines the dynamics in the absence of noise; ${\bf K}={\bf 0}$  at the stable state positions $\qb_{A_1}$, $\qb_{A_2}$  and at the saddle point $\qb_S$. We assume that  $\qb_S$ lies on a smooth hypersurface that separates the basins of attraction of states $A_1$ and $A_2$, cf. Fig.~\ref{fig:phase_portrait}. The function $\fb(t)$ in Eq. (\ref{eq:Langevin}) is white Gaussian noise; the results can be also extended to colored noise. The noise intensity $D$ is assumed small. The dependence of the switching rates $W_{nm}$ on $D$ is given by the activation law, $\log W_{nm}\propto D^{-1}$ \cite{Freidlin_book,Dykman1979a,Graham1984,Luciani1987,Bray1989,Dykman1990,Maier1993}. This is also the case for noise-driven continuous systems, cf. Ref.~\onlinecite{Smelyanskiy1997c,Maier2001,Fogedby2004} and papers cited therein. There exists extensive literature on numerical calculations of the switching rate and switching paths, cf. Ref.~\onlinecite{Eastman2001,Bolhuis2002,E2004,Zuckerman2004,Branduardi2007} and papers cited therein.

In the model (\ref{eq:Langevin}), the characteristic relaxation time $t_r$ and the characteristic diffusion length $l_D$ are
\begin{equation}
\label{eq:t_r_and_l_D}
t_r = \max_k |{\rm Re}\, \lambda_k|^{-1}, \qquad l_D=(Dt_r)^{1/2},
\end{equation}
where $\lambda_k$ are the eigenvalues of the matrix $\partial K_m/\partial q_n$ calculated at $\qb_{A_1},\qb_{A_2}$ and $\qb_S$.

For a white-noise driven system (\ref{eq:Langevin}), the three-time probability distribution $\rho(\qb_f,t_f;\qb,t|\qb_0,t_0)$ in Eq.~(\ref{eq:1}) can be written as a product of two-time transition probability densities,
\begin{equation}
\label{eq:transition_probability_decoupling}
 \rho(\qb_f,t_f;\qb,t|\qb_0,t_0)=\rho(\qb_f,t_f|\qb,t)
\rho(\qb,t|\qb_0,t_0),
\end{equation}
which simplifies further analysis. The analysis is done separately for the case where the observation point $\qb$ lies within the attraction basins of the initially empty attractor $A_2$ and the initially occupied attractor $A_1$.

\subsection{Switching probability distribution in the initially unoccupied basin of attraction}
\label{subsec:downhill}

We start with the case where the observation point $\qb$ lies in the basin of attraction of the initially empty state $A_2$ far from the stationary states, $|\qb - \qb_S|, |\qb-\qb_{A_{1,2}}| \gg l_D$.  For weak noise intensity, the system found at such $\qb$ will most likely approach $\qb_{A_2}$ over time $\tr$ moving close to the noise-free trajectory $\dot \qb = {\bf K}$ and will then fluctuate about $\qb_{A_2}$. Therefore, for $\qb_f$ not far from the attractor $A_2$, i.e., $|\qb_f-\qb_{A_2}|\lesssim l_D$,  we have
$\rho(\qb_f,t_f|\qb,t) \approx \rho_2(\qb_f).$
Here, $\rho_2(\qb_f)$ is the stationary probability distribution in the attraction basin of $A_2$ in the neglect of $A_2\to A_1$ switching. In its central part it has the form of a normalized Gaussian peak centered at $\qb_{A_2}$, with typical size $l_D$. Then, from Eq. (\ref{eq:1})
\[\p12 =\rho(\qb,t|\qb_0,t_0).\]

The analysis of the transition probability density $\rho(\qb,t|\qb_0,t_0)$ in this expression is simplified by two observations. First, for time $t$ in the range $W_{12}^{-1}\gg t-t_0 \gg \tilde{t_r}$, there is a probability current from attractor $A_1$ to $A_2$. This current gives the switching rate $W_{12}$, as found by Kramers \cite{Kramers1940}. The current density far from $A_2$, i.e.,  for $|\qb-\qb_{A_2}| \gg l_D$, is independent of time and is determined by the stationary Fokker-Planck equation
\begin{equation}
\label{eq:FPE}
 [-\partial_{\qb}{\bf K}+D\partial_{\qb}^2]\rho(\qb,t|\qb_0,t_0) = 0.
\end{equation}

The second observation is that, for both white and colored Gaussian noise, in switching the system is most likely to go close to the saddle point \cite{Dykman1990,Maier1997}. Having passed through the region near the saddle point the system moves close to the deterministic downhill trajectory from ${\cal S}$ to $A_2$, cf. Fig.~\ref{fig:phase_portrait}. This trajectory is described by equation $\dot\qb = {\bf K}$ and  gives the MPSP in the basin of attraction of $A_2$. We are interested in finding $\rho(\qb,t|\qb_0,t_0)$ for $\qb$ close to this trajectory. The broadening of the distribution is due to diffusion, which should generally make it  Gaussian in the transverse direction \cite{Ludwig1975}.

We parameterize the deterministic section of the MPSP by its length $\xi_{\parallel}$ counted off from $\qb_S$ and introduce a unit vector $\hat{\bm\xi}_{\parallel}$ along the vector ${\bf K}$ on the MPSP and $N-1$ vectors ${\bm\xi}_{\perp}$ perpendicular to it. The velocity on the MPSP is $v\equiv v(\xi_{\parallel})= K(\xi_{\parallel},{\bm\xi}_{\perp}={\bf 0})$. Of interest for our analysis are the values of $|\xib_{\perp}|$ of the order of the width of the path distribution transverse to the MPSP, which is given by the diffusion length, i.e., $|\xib_{\perp}|\lesssim l_D$. We assume $|\xib_{\perp}|$ to be small compared to the radius of curvature $|d\hat\xib_{\parallel}/d\xi_{\parallel}|^{-1}$.

Equation (\ref{eq:FPE}) can be solved near the MPSP by changing to variables $\xi_{\parallel}, \xib_{\perp}$, expanding ${\bf K}$ to first order in $\xib_{\perp}$, and replacing $\partial_{\qb}^2\to \partial_{\xib_{\perp}}^2$. One then obtains for $\rho(\qb,t|\qb_0,t_0)=\p12$ expression (\ref{eq:FPE}), with matrix $\hat Q$ given by equation
\begin{equation}
\label{eq:U-matrix}
 v\frac{d\hat{\rm Q}}{d\xi_{\parallel}}+\hat \kappa^{\dagger}\hat{\rm Q} +
 \hat{\rm Q}\hat\kappa + 2\hat{\rm Q}^2D=0.
 \end{equation}
Here, $\hat\kappa_{\mu\nu}=\partial K_{\mu}/\partial\xi_{\perp\,\nu}$, with the derivatives evaluated for ${\bm\xi}_{\perp}={\bf 0}$; the subscripts $\mu,\nu=1,\ldots,N-1$ enumerate the components of $\bm\xi_{\perp}$ and the transverse components of ${\bf K}$ in the co-moving frame. Equation (\ref{eq:U-matrix}) can be reduced to a linear equation for $\hat {\rm Q}^{-1}$. From Eq.~(\ref{eq:U-matrix}), the matrix elements ${\rm Q}_{\mu\nu}\propto 1/D$. Therefore the width of the switching probability distribution (2) is $\propto l_D$, as expected from qualitative arguments.

\subsection{Switching probability distribution in the initially occupied basin of attraction}

The case where the observation point $\qb$ lies in the basin of attraction of the initially occupied state $A_1$ is somewhat more complicated. Here, too, the two probability densities in the right-hand side of  Eq.~(\ref{eq:transition_probability_decoupling}) are independent of time $t$ for $|\qb-\qb_{A_{1,2}}|, |\qb-\qb_S|\gg l_D$. But in contrast to the previously studied region, none of them is known in advance. They have to be found from the Fokker-Planck equation (\ref{eq:FPE}) for $\rho(\qb,t|\qb_0,t_0)$ and the backward equation for $\rho(\qb_f,t_f|\qb,t)$,
\begin{equation}
\label{eq:Chapman-Kolmogorov}
 \left({\bf K}\partial_{\qb} +
 D\partial^2_{\qb}\right)\rho(\qb_f,t_f|\qb,t)=0.
 \end{equation}

We seek the solutions of Eqs.~(\ref{eq:FPE}) and (\ref{eq:Chapman-Kolmogorov}) in the eikonal form,
\begin{eqnarray}
\label{eq:eikonal_general}
& \rho(\qb,t|\qb_0,t_0)=\exp[-S_{\rm F}(\qb)/D],&\nonumber\\
 &\rho(\qb_f,t_f|\qb,t)=\exp[S_{\rm B}(\qb)/D]\rho_2(\qb_f).&
 \end{eqnarray}
The functions $S_{\rm F}$ and $S_{\rm B}$ can be written as power series in the noise intensity $D$, with $S_{\rm F, \,B}= S_{\rm F, \,B}^{(0)} + D S_{\rm F, \,B}^{(1)} +\ldots$. To the lowest order in $D$ we have
\begin{eqnarray}
\label{eq:Hamilton-Jacobi}
 H\left(\qb,\,\partial_{\qb}S_{{\rm F},\,{\rm B}}^{(0)}\right)=0,\qquad
 H(\qb,\pb)= \pb^2 + \pb{\bf K}(\qb).
\end{eqnarray}

Equation (\ref{eq:Hamilton-Jacobi}) has the form of a Hamilton-Jacobi equation for an auxiliary particle with coordinate $\qb$ and momentum $\pb$. This particle moves with energy $H=0$.   The functions $S_{\rm F,\,B}^{(0)}(\qb)$ are mechanical actions. Subscript ${\rm F}$ refers to motion of the auxiliary particle to point $\qb$ from the vicinity of $A_1$, as it is clear from Eq.~(\ref{eq:eikonal_general}). From Eq.~(\ref{eq:time_range}), this motion takes time that largely exceeds $t_r$. Using condition $H=0$, one can therefore associate $S_{\rm F}^{(0)}(\qb)$ with the mechanical action for reaching $\qb$ from $\qb_{A_1}$; the motion formally starts at $t\to -\infty$ from $\qb_{A_1}$, with momentum $\pb = {\bf 0}$ \cite{Freidlin_book}.

Subscript ${\rm B}$ in Eq.~(\ref{eq:Hamilton-Jacobi}) refers to the auxiliary Hamiltonian particle that moves from $\qb$ further away from attractor $A_1$. In this motion the original system goes close to the saddle point, and so should the auxiliary particle, too. The perturbation theory that underlies Eq.~(\ref{eq:Hamilton-Jacobi}) applies where the particle is approaching the saddle point, but has not gone beyond it. Indeed, for $H=0$ the particle approaches the saddle point asymptotically, for infinite time. Therefore $S_{\rm F}^{(0)}$ is the mechanical action for reaching $\qb_{\cal S}$ from $\qb$.

From Eqs.~(\ref{eq:transition_probability_decoupling}), (\ref{eq:eikonal_general}), the MPSP inside the basin of attraction of the initially occupied state corresponds to the maximum of $S_{{\rm B}}^{(0)}(\qb)-S_{\rm F}^{(0)}(\qb)$ and thus is determined by equation
\begin{equation}
\label{eq:extremum_on_MPSP}
 \partial_{\qb}S_{\rm F}^{(0)} = \partial_{\qb}S_{\rm B}^{(0)}.
\end{equation}
%(10)
The MPSP is thus given by the heteroclinic Hamiltonian trajectory that goes from the state $(\qb_{A_1},\pb = {\bf 0})$ to $(\qb_{\cal S}, \pb ={\bf 0})$.

To find $S_{{\rm F},\,{\rm B}}^{(0)}(\qb)$ close to the MPSP it is convenient to switch to a co-moving frame on the MPSP $(\xib_{\parallel}, \xib_{\perp})$. From Hamiltonian (\ref{eq:Hamilton-Jacobi}), the longitudinal direction $\hat\xib_{\parallel}$ and the velocity on the MPSP are given by expression
\begin{eqnarray}
\label{eq:MPSP_condition}
 2\partial_{\qb}S_{\rm F}^{(0)}(\qb) +{\bf K}(\qb) =
 v(\xi_{\parallel})\hat\xib_{\parallel},
\end{eqnarray}
%(11)
where the left-hand side is calculated for $\xib_{\perp}={\bf 0}$. [Eq.~(\ref{eq:MPSP_condition}) applies also if we use $S_{\rm B}^{(0)}$ instead of $S_{\rm F}^{(0)}$]. Note that the MPSP direction $\hat\xib_{\parallel}$ is not along the velocity of the original system in the absence of noise ${\bf K}$, in the general case of a system lacking detailed balance.

Close to the MPSP we can expand $S_{{\rm F},\,{\rm B}}^{(0)}$ and ${\bf K}$ in $\xib_{\perp}$. From Eqs. (\ref{eq:1}), (\ref{eq:eikonal_general}) and (\ref{eq:extremum_on_MPSP}), $\p12\propto \exp(-\xib_{\perp}\hat{\rm Q}\xib_{\perp}/2)$, as in Eq.~(\ref{eq:2}). The matrix $\hat{\rm Q}$ is expressed in terms of the actions $S_{\rm F,\,B}^{(0)}$ close to the MPSP as
\begin{eqnarray}
\label{eq:difference_matrix}
 && \hat{\rm Q} = \hat{\rm Q}_{\rm F}-\hat {\rm Q}_{\rm B},\nonumber\\
 && (\hat{\rm Q}_{\rm F,\,B})_{\mu\nu}=D^{-1}\partial^2 S_{\rm F,\,B}^{(0)}/\partial\xi_{\perp\mu}\partial\xi_{\perp\nu},
 \end{eqnarray}
%(12)
with the derivatives calculated for $\xib_{\perp} = {\bf 0}$. From the condition that $\p12$ be maximal on the MPSP it follows that matrix $\hat{\rm Q}$ is positive definite.

\subsubsection{The prefactor}

Interestingly, the prefactor in $\p12$ can be expressed explicitly in terms of the velocity $v(\xi_{\parallel})$ and the matrix $\hat{\rm Q}$. Formally, the prefactor is determined by the terms $S_{\rm F,\,B}^{(1)}$ in Eq.~(\ref{eq:eikonal_general}). The equations for them follow from Eqs.~(\ref{eq:FPE}) and (\ref{eq:Chapman-Kolmogorov}),
\begin{eqnarray}
\label{eq:S_F_general}
& \left(2\partial_{\qb}S_{\rm F}^{(0)} + {\bf K}\right)\partial_{\qb}S_{\rm F}^{(1)} - \partial_{\qb}{\bf K}
 -\partial_{\qb}^2S_{\rm F}^{(0)}=0,&\nonumber\\
 & \left(2\partial_{\qb}S_{\rm B}^{(0)} + {\bf
 K}\right)\partial_{\qb}S_{\rm B}^{(1)} +\partial_{\qb}^2S_{\rm B}^{(0)}=0. &
\end{eqnarray}
From Eqs.~(\ref{eq:MPSP_condition}), (\ref{eq:S_F_general}), to leading order in $\xib_{\perp}$ we have
\begin{eqnarray}
\label{eq:S_F_near_MPSP_general}
 & v\partial_{\xi_{\parallel}}S^{(1)}- \partial_{\xi_{\parallel}}v -
 \Tr\left[\hat\kappa + D\left(\hat{\rm Q}_{\rm F} + \hat{\rm Q}_{\rm B}\right)\right]=0, & \nonumber\\
 &  S^{(1)}=S_{\rm F}^{(1)}-S_{\rm B}^{(1)}, &
 \end{eqnarray}
%(14)
where, as before, $\kappa_{\mu\nu}=\partial K_{\mu}/\partial\xi_{\perp\nu}$ with the derivative calculated for $\xib_{\perp}={\bf 0}$.

On the other hand, by expanding in Hamilton-Jacobi equations
(\ref{eq:Hamilton-Jacobi}) for $S_{\rm F,\,B}^{(0)}$ near the MPSP to second order in $\xib_{\perp}$ and taking into account the relation between the derivatives of $S_{\rm F}^{(0)}$ and $S_{B}^{(0)}$ on the MPSP (\ref{eq:extremum_on_MPSP}), (\ref{eq:MPSP_condition}) we obtain an important relation
\begin{eqnarray*}
\label{eq:matrix_difference}
 v\partial_{\xi_{\parallel}}\hat{\rm Q} + 2D\left(\hat{\rm Q}_{\rm F}^2-\hat{\rm Q}_{\rm B}^2\right)+ \hat\kappa^{\dagger}\hat{\rm Q} + \hat{\rm Q}\hat\kappa=0.
\label{eq:important_relation}
\end{eqnarray*}
From this equation
\[\Tr\left[\hat\kappa + D\left(\hat{\rm Q}_{\rm F} + \hat{\rm Q}_{\rm B}\right)\right] =-\frac{1}{2}v\partial_{\xi_{\parallel}}\Tr\log\hat{\rm
 Q}.\]

By substituting this relation into Eq.~(\ref{eq:S_F_near_MPSP_general}) we obtain
\begin{equation}
\label{eq:S_F_answer}
 S^{(1)}(\xi_{\parallel})= \log
 v(\xi_{\parallel})-\frac{1}{2}\Tr\log\hat{\rm Q}(\xi_{\parallel}) + \log C_1,
\end{equation}
where we explicitly indicate that $S^{(1)}, \, v$, and $\hat{\rm Q}$ are functions of the distance, $\xi_{\parallel}$, along the MPSP;  $C_1$ is a constant of integration.

Equations (\ref{eq:1}), (\ref{eq:eikonal_general}), (\ref{eq:difference_matrix}), and (\ref{eq:S_F_answer}) lead to expression (\ref{eq:2}) for the switching probability distribution. Note that, from Eq.~(\ref{eq:difference_matrix}), inside the initially occupied basin of attraction the width of the peak of the distribution transverse to the MPSP is $\sim l_D\propto D^{1/2}$. The distribution describes a stationary probability current. This current is the same in the basins of attraction of states $A_1$ and $A_2$. In obtaining Eq.~(\ref{eq:2}) from Eq.~(\ref{eq:S_F_answer}) we found $C_1$ from the condition $v(\xi_{\parallel})\int d\xib_{\perp}\p12 =W_{12}$.

From conservation of the stationary probability current it follows that the distribution $\p12$ should sharply increase near the saddle point. Indeed, the velocity $v(\xi_{\parallel})= 0$ for $\qb = \qb_S$. The current close to $\qb_S$ is due to diffusion. In the general case of nonequilibrium systems the shape of the switching probability distribution near the saddle point is complicated; its analysis is beyond the scope of this paper.

\subsection{Switching probability distribution for systems with detailed balance}

An explicit solution for $\p12$ near the saddle point can be obtained for systems with a gradient force ${\bf K}=-\partial_{\qb}U(\qb)$. Such systems have detailed balance. The uphill section of the MPSP is literally the uphill path that goes from the local minimum of the potential $U(\qb)$ at $A_1$ to the saddle ${\cal S}$ and is given by equation $\dot\qb = \partial_{\qb}U(\qb)$ \cite{Onsager1953} [this can be seen from Eq.~(\ref{eq:Hamilton-Jacobi})]. In contrast to systems without detailed balance \cite{Dykman1994c,Maier1997}, for smooth $U(\qb)$ the MPSP near the saddle point is described by an analytic function of coordinates and $\hat\xib_{\parallel}$ is perpendicular to the separating hypersurface.

The quasistationary solution of the forward Fokker-Planck equation (\ref{eq:FPE}) near a saddle point has been known since the work of Kramers \cite{Kramers1940} and Landauer and Swanson \cite{Landauer1961}. The backward equation (\ref{eq:Chapman-Kolmogorov}) can be solved similarly by expanding ${\bf K}$ to first order in $\qb-\qb_S$ and by using the condition that deep inside the basin of attraction of the initially empty state $A_2$ we have $\rho(\qb_f,t_f|\qb,t)\approx \rho_2(\qb_f)$. The solution has the form
\begin{eqnarray}
\label{eq:backward_near_saddle}
&\rho(\qb_f,t_f|\qb,t)\approx \frac{1}{2}\rho_2(\qb_f)\left[1+{\rm erf} (\tilde\xi_{\parallel})\right],&\nonumber\\
& \tilde\xi_{\parallel}=
(\lambda_{\parallel}/2D)^{1/2}(\xi_{\parallel}-\xi_{\parallel\,{\cal S}}). &
\end{eqnarray}
Here, ${\rm erf}(x)$ is the error function, $\xi_{\parallel\,{\cal S}}$ is the position of the saddle point on the MPSP, and $\lambda_{\parallel}$ is the curvature of the potential $U(\qb)$ at the saddle point in the steepest descent direction $\hat\xib_{\parallel}$, $U(\qb) \approx -\lambda_{\parallel}(\xi_{\parallel}-\xi_{\parallel\,{\cal S}})^2/2$ for $\xib_{\perp}={\bf 0}$ and small $|\xi_{\parallel}-\xi_{\parallel\,{\cal S}}|$.

Equation (\ref{eq:backward_near_saddle}) combined with the results \cite{Kramers1940,Landauer1961} give expression~(\ref{eq:2}) for $\p12$ near $\qb_S$ provided one replaces in this expression
\begin{eqnarray}
\label{eq:replacement_near_saddle}
v^{-1}(\xi_{\parallel}) \to
 (\pi/8\lambda_{\parallel}D)^{1/2}\exp({\tilde\xi_{\parallel}}^2)
 \left[1-{\rm erf}^2(\tilde\xi_{\parallel})\right],
 \end{eqnarray}
%(16)

Equation (\ref{eq:replacement_near_saddle}) goes over into  $v^{-1}(\xi_{\parallel})$ for $|\xi_{\parallel}-\xi_{\parallel\,{\cal S}}| \gg l_D$. In the opposite limit, that is very close to the saddle point, it shows that $v^{-1}$ is replaced by a factor $(\pi/8\lambda_{\parallel}D)^{1/2}\sim t_r/l_D$. This demonstrates that the distribution $\p12$ does not diverge at the saddle point, but it contains a large factor $D^{-1/2}$.

\section{MICROMECHANICAL TORSIONAL OSCILLATOR}
\label{sec:system}

\subsection{Device characteristics}

\begin{figure}[h]
\includegraphics[width=3.0in]{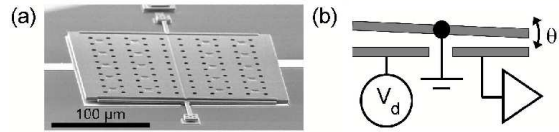}
\caption{Micromechanical torsional oscillator used for studying the switching path distribution. (a) Scanning electron micrograph. (b) Cross-sectional schematic. The angle $\theta$ of the movable plate is controlled by the voltage applied to one of the fixed electrodes. Oscillation of the plate is detected using the other electrode.}
\label{fig:torsional_oscillator}
\end{figure}

We measure the switching probability distribution using a high-$Q$ micro-electromechanical torsional oscillator ($Q = 9966$) driven into parametric resonance. The oscillator is shown in Fig.~\ref{fig:torsional_oscillator}. It consists of a movable, highly-doped polysilicon plate (200~$\mu$m $\times$ 200~$\mu$m $\times$ 3.5~$\mu$m) suspended by two torsional rods (4~$\mu$m $\times$ 2~$\mu$m $\times$ 36~$\mu$m, spring constant = $3.96 \times 10^{-8} $~Nm). There are two fixed electrodes on the substrate, one on each side of the torsional rod. The 2~$\mu$m gap underneath the movable plate is created by etching away a sacrificial silicon oxide layer.

Torsional oscillations of the movable top plate are excited by applying a driving voltage $V_d = V_{\rm dc} + V_{\rm ac} \cos{(\omega t)} + V_{\rm noise}(t)$   to one of the lower electrodes while the top plate remains electrically grounded. The driving frequency $\omega = 2\omega_0 + \varepsilon$  is close to twice the natural frequency $\omega_0$. The dc voltage $V_{\rm dc}$ (1~V) is much larger than the amplitude $V_{\rm ac}$ (141 mV) of sinusoidal modulation and the random noise voltage $V_{\rm noise}$.

Because of the applied voltage, the top plate is subjected to an electrostatic torque $\tau = (dC/d\theta)V_d^2/2$ in addition to the restoring torque of the springs:
\begin{equation}
\label{eq:restoring_torque}
 \ddot \theta + 2\Gamma \dot \theta + \omega_0^2 \theta = \frac{\tau}{I},
 \end{equation}
%(17)
where  $\theta$ is the rotation angle (see Fig.~\ref{fig:torsional_oscillator}), $\Gamma$  is the damping constant, and $I$ is the moment of inertia of the plate. The time-independent component of the torque $G=-C'(\theta_0)V_{\rm dc}^2/2$ leads to a shift of the equilibrium position of $\theta$ to $\theta_0$. In what follows we count $\theta$ off from $\theta_0$.

The expansion of the torque in $\theta$ to first order in $V_{\rm ac}, V_{\rm noise}$ has the form
\begin{eqnarray}
\label{eq:restoring_torque_2}
I^{-1}\tau &\approx &-\left[k_0 + k_e\cos{(\omega t)}\right]\theta  - \alpha\theta^2 -\beta\theta^3  \nonumber\\
 && + F\cos{(\omega t)}+N(t).
 \end{eqnarray}
Here, $k_0=-C''(\theta_0)V_{\rm dc}^2/2I$, $\alpha=-C^{(3)}(\theta_0)V_{\rm dc}^2/4I$, and $\beta=-C^{(4)}(\theta_0)V_{\rm dc}^2/12I$ are the linear and nonlinear coefficients of the electrostatic torque; because the oscillation amplitude remains small, we disregard terms of higher order in $\theta$.

The ac voltage leads to a time-dependent additive torque with amplitude $F=C'(\theta_0)V_{\rm dc}V_{\rm ac}(t)I$ and also to modulation of the spring constant $\propto k_e$, with $k_e=-C''(\theta_0)V_{\rm dc}V_{\rm ac}/I$. Since the $Q$-factor of the oscillator is high, the response at $\omega \sim 2\omega_0$ is negligible. Therefore the major effect of the ac voltage is the parametric modulation $k_e\theta\cos{(\omega t)}$. The renormalization of this term by the nonlinear terms in Eq.~(\ref{eq:restoring_torque_2}) [for example, $\propto \alpha\theta \omega_0^{-2}F\cos\omega t$] is small for our device and is disregarded. The term $N(t)=C'(\theta_0)V_{\rm dc}V_{\rm noise}(t)/I$ represents zero-mean noise in the driving torque. This noise is Gaussian and its spectrum is flat in a broad frequency range that goes from zero frequency far beyond $\omega_0$; therefore for our purpose it can be assumed white, with intensity $\tilde D$ given by $\left\langle N(t)N(t')\right\rangle = 2\tilde D\delta(t-t')$.

With Eq.~(\ref{eq:restoring_torque_2}), the equation of motion for the angle $\theta$ counted off from $\theta_0$ becomes
\begin{equation}
\label{eq:reduced_eq_of_motion}
 \ddot\theta +2\Gamma\dot\theta + \left[\omega_1^2 + k_e \cos{(\omega t)}\right]\theta +\alpha\theta^2 +\beta\theta^3 = N(t),
 \end{equation}
where $\omega_1^2 = \omega_0^2 + k_0$ (in our device the difference $\left|\omega_0-\omega_1\right|$ is small, $< 0.001\omega_0$).

Torsional oscillations of the top plate are detected capacitively by the other electrode. This electrode is connected to a dc voltage source through a large resistor. A high electron mobility transistor is placed in close proximity to the device to measure the oscillating charge on the detection electrode induced by motion of the top plate. The output of the transistor is connected to a lock-in amplifier referenced at half the driving frequency $\omega$. For the chosen time constant of 300 $\mu s$, the measurement uncertainty is $\sim$ 80 $\mu$rad, about 0.6\% of the full scale in Fig.~\ref{fig:1} and much smaller than the width of the path distributions. The oscillation amplitudes in-phase (X) and out-of-phase (Y) with the reference frequency were recorded every 2 ms. All measurements were performed at 77 K and $<10^{-6}$ torr.

\subsection{Transformation to slow variables and parametric resonance}
\label{subsec:slow_variables}

Since the oscillator is strongly underdamped $(\Gamma/\omega_1 \sim 10^{-4} )$ and the modulation is almost resonant $(\left| \omega - 2\omega_1 \right| \ll \omega)$, we analyze the motion of the oscillator in the rotating frame, with slow dimensionless variables $q_1$ and $q_2$ and dimensionless time $t \to k_e t/2\omega$ (note that even though the oscillator has one degree of freedom, its motion is characterized by two dynamical variables). In our micromechanical device the characteristic renormalized parameter of cubic nonlinearity $\gamma = \beta - (10\alpha^2/9\omega_1^2)$ is negative. In this case it is convenient to introduce the slow variables as
\begin{eqnarray}
\label{eq:motion_rotating_frame}
 \theta (t) = \left(\frac{2k_e}{3\vert\gamma\vert}\right)^{1/2}\left[q_1\cos\left(\frac{\omega t}{2}\right) - q_2\sin\left(\frac{\omega t}{2}\right)\right],\nonumber\\
 \frac{d\theta}{dt}  = -\left(\frac{\omega^2 k_e}{6\vert\gamma\vert}\right)^{1/2}\left[q_1\sin\left(\frac{\omega t}{2}\right) + q_2\cos\left(\frac{\omega t}{2}\right)\right].
 \end{eqnarray}
%(20)
The variables $q_1$ and $q_2$ are interchanged here compared to Ref.~\onlinecite{Ryvkine2006a}, which referred to the case $\gamma>0$.

The quadratures $q_1$  and $q_2$  are directly proportional to the signal components $X$ and $Y$ measured with the lock-in amplifier, with the proportionality constant $E$ determined by the measuring apparatus,
\begin{equation}
\label{eq:proportionality_C}
 q_1 = EX, \qquad q_2= EY.
 \end{equation}

Substituting Eq.~(\ref{eq:motion_rotating_frame}) into Eq.~(\ref{eq:reduced_eq_of_motion}) and neglecting fast oscillating terms, we can write the equations of motion for $\qb=(q_1,q_2)$ in the form (\ref{eq:Langevin}). The function $\bf K$ in dimensionless time is given by
\begin{equation}
\label{eq:K}
 \bf K =-\zeta^{-1}\qb + \hat\varepsilon \nabla g.
 \end{equation}
 %(22)
Here $\zeta = k_e/2\omega \Gamma$, $\mu = \omega(2\omega_1-\omega)/k_e$, and $ g =q^4/4 - (1+\mu)q_1^2/2+(1-\mu)q_2^2/2$, where $\hat\varepsilon$ is the permutation tensor (the parameter $\mu$ is defined with the opposite sign compared to Ref.~\onlinecite{Ryvkine2006a}, again because $\gamma$ has the opposite sign for our system). Equation $\dot \qb = {\bf K}$ gives the downhill section of the MPSP of the oscillator. The uphill section of the MPSP can be calculated by solving the Hamiltonian equations of motion that follow from Eq.~(\ref{eq:Hamilton-Jacobi}).

\subsection{Determination of device parameters}

We first consider motion of the device in the absence of fluctuations. When the amplitude of the spring modulation is sufficiently strong $(\zeta >1)$, the oscillator response exhibits period doubling \cite{LL_Mechanics2004}. Oscillations are induced at half the modulation frequency in a range close to $\omega_1$. Between the two bifurcation frequencies $\omega_{b1}$  ($\approx 139318.11$~rad/s) and $\omega_{b2}$  ($\approx 139384.74$~rad/s) there exists two stable states of oscillations at frequency $\omega /2$. They differ in phase by $\pi$ but have identical amplitude. Both states are stable solutions of Eq.~(\ref{eq:reduced_eq_of_motion}). Their basins of attraction in the rotating frame are separated by a separatrix that goes through the unstable stationary state, which in the laboratory frame has zero vibration amplitude at frequency $\omega/2$. The phase portrait in the rotating frame is illustrated in Fig.~\ref{fig:phase_portrait}. The driving frequency is chosen to be 278639.16 rad/s for measurement of the switching path distribution. 

We note that parametric resonance in nano- and micro-electro-mechanical systems has attracted considerable attention \cite{Rugar1991,Turner1998,Buks2002,Lifshitz2003,Cleland2005,Mahboob2008}. Since here we are interested in the studies of the principal features of noise-induced switching, we chose the simplest nontrivial regime where the system has only two stable states, which occurs for $\omega_{b1}<\omega/2 < \omega_{b2}$. The modulation frequency  $\omega$ is chosen to be close to $2 \omega_{b1}$ so that the motion in the rotating frame is underdamped, which is advantageous for studying a generic feature of fluctuations in systems far from thermal equilibrium, the breaking of time reversal symmetry.

%%%%%%%%%%%%%%%%%%%%%
\begin{figure}[h]
\includegraphics[width=3.0in]{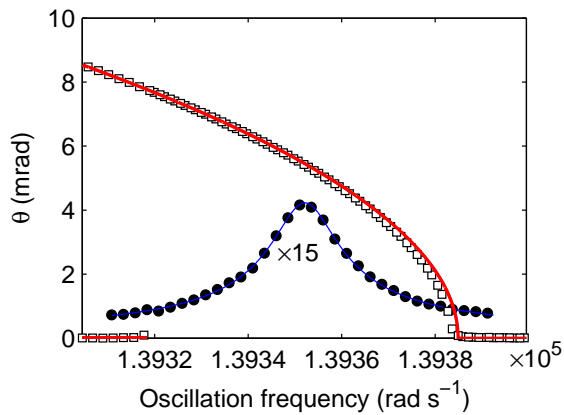}
\caption{\label{fig:2} (Color online) Harmonic and parametric resonances of the micromechanical torsional oscillator. For resonant driving (solid circles), the oscillation amplitude is plotted as a function of the oscillation frequency. The thin line is a fit to the harmonic oscillator response. It gives device parameters $\Gamma$ and $\omega_1$. For parametric resonance (hollow squares), the driving frequency is twice the oscillation frequency. The fit (thick line) yields $\kappa_{\rm nonlinear}$ and the effective parametric modulation amplitude $k_e$.}
 \end{figure}
 %%%%%%%%%%%%%%%%%%%%%%

Calculation of the MPSP requires a number of device parameters including $\Gamma$, $\omega_1$, the parametric modulation amplitude $k_e$, and the nonlinear constant $\kappa_{\rm nonlinear} = 3\gamma/8\omega_1$ \cite{Ryvkine2006a}. These parameters are obtained from the linear and nonlinear responses of the device. When the device is resonantly driven with small amplitude at frequency close to $\omega_1$, it responds as a harmonic oscillator. From the resonance line shape (Fig. \ref{fig:2}), $\Gamma$ and $\omega_1$ are determined to be 6.99~rad/s and 139352.118~rad/s respectively. The remaining two parameters are extracted from the parametric resonance of the oscillator for $\omega$ close to $2\omega_1$. Specifically, the parametric modulation amplitude $k_e$ is determined from the bifurcation frequencies $\omega_{b1,2} = 2\omega_1 \mp \omega_p$, where $\omega_p = (k_e^2-(4\omega_1\Gamma)^2)^{1/2}/2\omega_1$. This gives $k_e= 1.94\times 10^{-7}\;\rm s^{-2}$. The nonlinear parameter $\kappa_{\rm nonlinear}$(1.08$\times10^6\rm\;s^{-1}$) is obtained from the proportionality constant between the square of the parametric oscillation amplitude $\theta_A$ and the detuning from bifurcation frequency close to the bifurcation frequency seen in Fig.~\ref{fig:2},
\begin{equation}
\label{eq:proportionality_const}
 \theta_A^2 = (\omega - \omega_{b2})/2\kappa_{\rm nonlinear}.
 \end{equation}

Using these measured device parameters, the dimensionless constants contained in $\bf K$ in Eq.~(\ref{eq:K}) and in Eq.~(\ref{eq:proportionality_C}) can be directly calculated to be $E = 176.349$, $\zeta$ = 4.968, and $\mu$ =-0.9367. The theoretical optimal escape path in Fig.~\ref{fig:1} is calculated with the above parameter values. No adjustable parameters are used.

\section{SWITCHING PATH DISTRIBUTION: EXPERIMENT}
\label{sec:experiment}

\subsection{Measured switching path distribution}

When white noise is added to the excitation voltage, the system can occasionally overcome the activation barrier and switch from one stable state to the other. The noise intensity is chosen to ensure that the mean residence time in each state ($\sim$ 10 s) is much larger than the relaxation time ($t_r \sim$ 1 s) of the system. Transitions are identified when the oscillator begins in the vicinity of $A_1$  (within the left green circle $\Omega_1$  in Fig.~\ref{fig:1}a) and subsequently arrives at state $A_2$ (within the right green circle $\Omega_2$). Figure~\ref{fig:1} shows the switching probability distribution derived from more than 6500 transitions. While in each transition the system follows a different trajectory, the trajectories clearly lie within a narrow tube.

The maximum of the distribution gives the MPSP. In Fig.~\ref{fig:1}b, the location of this maximum is plotted on top of the MPSP obtained from theory. The oscillator is underdamped not only in the laboratory frame, but also in the rotating frame. Therefore both the uphill and downhill sections of the MPSP are spirals. On the uphill section, the MPSP emerges clockwise from $A_1$ and spirals toward the saddle point at the origin. Upon exit from the saddle point, it makes an angle and, on the downhill section, continues to spiral clockwise toward $A_2$.

There is excellent agreement between the measured peak in the probability distribution and the MPSP obtained from theory. There are no adjustable parameters since all device parameters are accurately determined from the harmonic and parametric resonances of the oscillator without noise in the excitation as described in the previous section.

Close to the stable states the peaks of the distribution at successive turns of the MPSP overlap, preventing the accurate determination of the MPSP. The plot in Figs.~\ref{fig:1}a and \ref{fig:1}b has excluded the portions of trajectories prior to escaping from the initial state $A_1$ and upon arriving at the final state $A_2$, which are bound by the two blue lines. Such cutoff also eliminates the large peaks of the distribution centered at $A_1$ and $A_2$, which arise because the oscillator spends most of its time fluctuating about $A_1$ and $A_2$. These peaks are not relevant to switching dynamics.

\begin{figure}[h]
\includegraphics[width=2.4in]{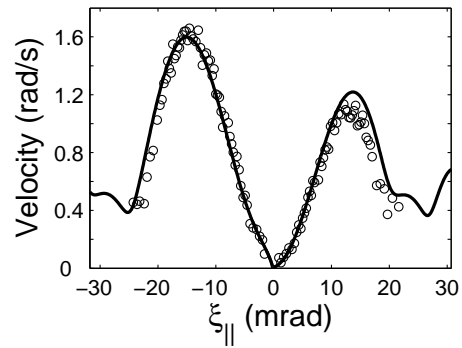}
\caption{Measured averaged velocity along the MPSP (circles) and the velocity predicted by theory (line). The velocity decreases to zero at the saddle point $(\rm \xi_{\parallel} = \bf 0)$.}
\label{fig:4}
\end{figure}

Figure~\ref{fig:4} compares the measured and predicted velocity along the MPSP. Here, again, the good agreement is demonstrated with no adjustable parameters. As expected, the measured velocity decreases near the saddle point, $\xi_{\parallel}=0$. However, it does not become equal to zero, in agreement with the argument that the total probability current remains constant. Motion near the saddle point is dominated by diffusion.

%%%%%%%%%%%%%%%%
%%%%%%%%%%%%%%

\subsection{Generic features of the switching path distribution}

\begin{figure}[h]
\includegraphics[width=3.4in]{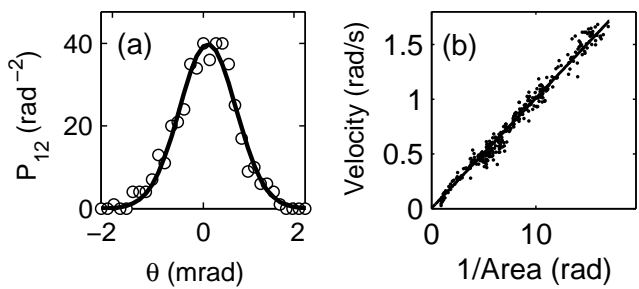}
\caption{(a) The cross-section along the purpleline in Fig.~\ref{fig:1} transverse to the MPSP. The solid line is a Gaussian fit. (b) Velocity on the MPSP vs. inverse area under cross-sections of the switching probability distribution. The solid line is a linear fit forced through the origin.}
\label{fig:3}
\end{figure}

The switching probability distribution in our multi-variable system displays important generic features. Figure~\ref{fig:3}a shows the distribution cross-section along the purple line transverse to the MPSP in Fig.~\ref{fig:1}b.  It is well-fitted by a Gaussian. Gaussian distributions with different height and area are observed also in other cross-sections except close to the saddle point. Figure~\ref{fig:3}b plots the area under the Gaussian distribution versus the reciprocal measured velocity on the MPSP, for different cross-sections. The linear dependence agrees with Eq.~(\ref{eq:2}) and indicates that the probability current from the initially occupied attractor to the empty one is constant. This current gives the switching rate $W_{12}$ \cite{Kramers1940}.

We find that the probability current concentrates within a narrow tube deep into the basins of attraction of $A_1$ and $A_2$. In the basin of attraction to $A_2$ but not too close to $A_2$, much of the probability distribution carries the switching current. However, the overall quasistationary probability distribution deep inside the basin of attraction of $A_1$ is largely associated with fluctuations about $A_1$  that do not lead to switching. The part of the distribution responsible for the switching current is an exponentially small fraction of the total distribution. Nevertheless our formulation makes it possible to single out and directly observe this fraction.

%%%%%%%%%%%%%%%%%%%%
%%%%%%%%%%%%%%%%%%

The slowing down near the saddle point shown in Fig.~\ref{fig:4} leads to strong broadening and increase in height of the switching probability distribution seen in Fig.~\ref{fig:1}a. Because motion near the saddle is diffusive, switching paths loose synchronization. In other words, the distribution of times spent by the system near the saddle point is comparatively broad. This is why it is advantageous to study the distribution of switching paths in the space of dynamical variables rather than in time.

\subsection{Lack of time reversal symmetry in a driven oscillator}

Another generic feature of the observed distribution is characteristic of systems far from thermal equilibrium. For equilibrium systems, the most probable fluctuational path uphill, i.e., from an attractor to the saddle point, is the time reversal of the fluctuation-free downhill path from the saddle point back to the attractor. More precisely, it corresponds to the change of the sign of dissipation term in the equation of motion \cite{Onsager1953,Machlup1953}, i.e., to replacing $\Gamma$  with $-\Gamma$  in Eq.~(\ref{eq:restoring_torque}). In overdamped equilibrium systems with detailed balance, these two paths coincide in space (but are opposite in direction).

\begin{figure}[h]
\includegraphics[width=2.0in, angle=-90]{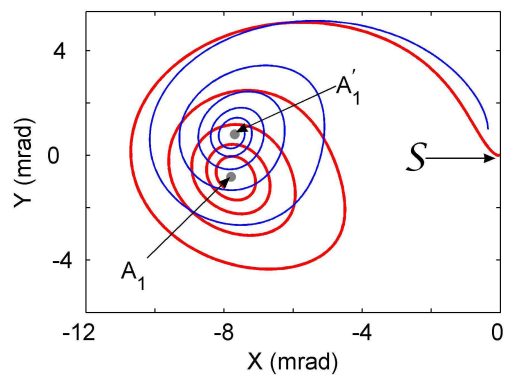}
\caption{(Color online) Comparison of the MPSP and the dissipation-reversed path. The section of the most probable switching path from $A_{1}$ to ${\cal S}$ is shown as a thick solid line. Upon changing the sign of dissipation, the attractor is shifted to a new location $A_{1}'$ and becomes a repeller. The fluctuation-free path with reversed dissipation from $A_{1}'$ to ${\cal S}$ is shown as the thin solid line.}
\label{fig:5}
\end{figure}

Our parametric oscillator is driven far from thermal equilibrium. Therefore the uphill section of the MPSP does not simply relate to the deterministic trajectory with reversed sign of dissipation. This section of the MPSP, i.e., the most probable fluctuational path from $A_1$ to the saddle point at the origin is plotted as the thick solid line in Fig.~\ref{fig:5}. Upon sign reversal of the dissipation, the attractor becomes a repeller, as in the case of systems in thermal equilibrium desribed earlier. However, in contrast to equilibrium systems, it is also shifted away from its original location (from $A_1$ to $A_1'$  in Fig. \ref{fig:5}). The dissipation-reversed path is shown as the thin solid line in Fig.~\ref{fig:5}. In addition, Fig.~\ref{fig:phase_portrait} allows one to compare the uphill section of the MPSP with the deterministic downhill path from ${\cal S}$ to $A_1$. Our data show that the uphill section of the MPSP, which is formed by fluctuations, the dissipation-reversed path, and the downhill noise-free path from the saddle to the stable state are all distinct. The time irreversibility of the switching paths is directly related to the lack of detailed balance of our driven oscillator, distinguishing it from bistable systems in thermal equilibrium.

%%%%%%%%%%%%%%%%%
%%%%%%%%%%%%%

\section{Conclusions}
\label{sec:conclusions}

In this paper we have studied the phase space distribution of paths followed in activated switching between coexisting stable states. The analysis refers to systems with several dynamical variables.  We introduced a quantity, the switching probability distribution, that gives the probability density of passing a given point in phase space during switching. The distribution is defined in a way that makes it experimentally accessible. No {\it a priori} knowledge of the system dynamics is required except the positions of the stable states, which can be immediately determined, since the system spends near these states most of the time.

The switching probability distribution was shown theoretically to have a shape of a narrow ridge in phase space. Far from the stationary states, the cross-section of the ridge is Gaussian. The maximum of the ridge lies on the most probable switching path (MPSP).

Experimental studies of the switching path distribution were done using a high-$Q$ micromechanical torsional oscillator. All parameters of the oscillator, including the nonlinearity constant, were directly measured. The oscillator was driven into parametric resonance, where it had two coexisting vibrational states that differ in phase. The paths followed in switching between these states were accumulated and their distribution in the space of the two dynamical variables (the oscillation quadratures) was obtained. It was found that the distribution has indeed the shape of a Gaussian ridge.

There is excellent agreement between the experimental and theoretical results, with no adjustable parameters. The measured maximum of the switching path distribution lies on top of the theoretically calculated MPSP. The measured velocity of motion along the MPSP as a function of the position on the MPSP also quantitatively agrees with the theory.  An important property of the path distribution is the total current conservation: the product of the velocity of motion along the MPSP and the cross-section area of the path distribution remains constant. This conservation of probability current was demonstrated experimentally. In addition, we observed, for the first time, that the lack of detailed balance leads to the difference between the uphill section of the MPSP and the noise-free path with reversed sign of dissipation.

The observation of the most probable switching path reported here provides, in some respects, an experimental basis for the broadly used concept of a reaction coordinate, which can be associated with the coordinate along this path. Our method does not rely on the specific model of the fluctuating system but only on the characteristics accessible to direct measurement. It applies to systems far from thermal equilibrium as well as to equilibrium systems. Measuring the switching trajectories can help to determine the model globally, far from the stable states. It can also provide an efficient way of controlling the switching rates by affecting the system locally on the most probable switching path.

This research was supported in part by NSF DMR-0645448 (HBC) and NSF PHY-0555346 and ARO W911NF-06-1-0324 (MID).

\bibliographystyle{apsrev}

\end{document}